\documentclass[pra,onecolumn]{revtex4-2}
\usepackage{graphicx}    
\usepackage{amsmath}
\usepackage{amsfonts}

\begin{document}

\title{Angular distributions and polarization correlations of the two-photon spherical states}
\author{Moorad Alexanian$^{1}$ and Vanik E. Mkrtchian$^{2}$}
\affiliation{$^{1}$Department of Physics and Physical Oceanography, University of North Carolina Wilmington, Wilmington, NC 28403-5606, USA\\
$^{2}$Institute for Physical Research, Armenian Academy of Sciences, Ashtarak 0203, Republic of Armenia}
\date{\today}

\begin{abstract}
\noindent \textbf{Abstract}. We have analyzed in detail the angular polarization properties in the center of mass reference frame of Landau's two-photon spherical states in momentum space. The angular distributions for fixed values of $J$ and $M$ do not depend on the parity but are defined by two different functions of the polar angle between the relative momentum and the quantization axes. The two-photon polarization density matrices are derived for each values of $J$, $M$, and $P$. The linear polarization correlations of individual photons are analyzed in detail. We find, besides the usual correlation laws for $J\geq 2$ in terms of $sin$ and $cos$ of the angle between the orientation of the analyzers, correlations in terms of the sum of the orientation angles of the analyzers.
\end{abstract}

\maketitle

\section{Introduction}

During the last eighty years researches in different areas of physics have found a vast number of applications of the two-photon (TP) states. In astrophysics, a first step was taken by Breit and Teller \cite{Br-Tel} who applied the calculations of Goeppert-Mayer \cite{GM} of the TP emission for the 2S-1S atomic transition showing that this process is the principal cause of the decay of interstellar hydrogen atoms and the continuum radiation from gaseous nebulae. Generalization of these calculations for highly excited levels of hydrogen atoms have been carried out \cite{Sun}. In addition, the processes of TP decay has been considered recently as a possible mechanism for the temperature and polarization anisotropies of the cosmic microwave background radiation \cite{Peeb}.

Two-photon emission is one of the principal tool in experiments testing violations of the Bell inequalities in atomic cascade and nuclei \cite {Asp,Klei,Gis}. These investigations test the fundamental question of quantum entanglement. Additionally, TP transitions have been proposed as a tool to measure weak interaction properties, viz., the problem of parity violations in atomic transitions \cite{Sch,Frat}. The previously indicated applications of the TP states concern interaction of the electromagnetic field in creation processes of TP by atomic systems, as well as, the properties of entanglement of many-particle states.

Landau \cite{Land} and Yang \cite{Yang} have analyzed symmetry properties of TP under spatial rotations and inversions in the center of mass reference frame of the system. Landau and Yang established selection rules for the TP decay of positronium and mesons. It is shown \cite{Land} that the TP annihilation of orthopositronium is forbidden since there is no state of the TP with angular momentum $J=1$ and the decay is primarily into three gamma photons. In addition, Yang \cite{Yang} also points out the kind of correlation that exists between the planes of polarizations of the TP emitted in the decay of the $\pi ^{0}$ meson.

The fact that two photons cannot have total angular momentum $J=1$ is a direct consequence of Bose-Einstein statistics and is referred to as the Landau-Yang theorem. The Landau-Yang theorem forbids two photons to participate in any process that would require them to be in a state with total angular momentum $J=1$. The connection of the Landau-Yang theorem with some specific selection rules for two-photon transitions in atoms was first emphasized \cite{DBD83,KEB09}, where an experimental limit for the violation of Bose-Einstein statistics was obtained. The Landau-Yang theorem for atomic transitions has been extended to three- and four-photon transitions \cite{ZSL15,ZSL17}.

Landau \cite{Land} introduces two-photon spherical states (TPSS) that are eigenfunctions of energy, parity $P$, angular momentum $J$, and its projection $M$ on the fixed quantization axis. Landau analyzes in detail the general transformation properties of TPSS wave functions in momentum space, in the center of mass reference frame. Table I indicates the number $N_{JM}^{(P )}$ of possible TPSS with given angular momentum $J$, projection $M$, and parity $P$.
\begin{table}[th]
\caption{All possible two-photon spherical states $|JMP\rangle$ ($k= 1, 2,\cdots)$}
\label{table:two}
\centering 
\begin{tabular}{ccc}
\hline\hline
$J$ &  $N_{JM}^{(+)}$ & $N_{JM}^{( -)} $\\[0.5ex] \hline
0 & $1$ & $1$ \\
1 & $-$ & $-$ \\
2$k$&2&1\\
2$k$+1 & 1 & $-$\\[1ex] \hline
\end{tabular}
\end{table}

The explicit expressions for the TPSS appears in the Wick-Jacob review paper \cite{Jacob} as an application of their helicity formalism. They first introduce TP spherical helicity states $\left\vert JM;\lambda \lambda ^{\prime }\right\rangle $ (see Chap. 2) as eigenfunctions of angular momentum $J$, and its projection $M$ and the total helicity of photons. They then show that each Landau TPSS can be represented as a sum or difference of two TP spherical helicity states with opposite total helicities \cite{Jacob, LLIV} (see Table II below).

The one-particle spherical states together with the theory of angular polarization correlations have been developed both in the classical \cite{Jack} and in the quantum theories \cite{Akh}.  The present article is devoted to analyzing the angular distribution and the polarization correlations in the TPSS.

The structure of this article is as follows. In Sec. II, after a short, preliminary consideration of the explicit form of each two-photon Landau spherical states, we analyze their angular distribution in momentum space relative to the quantization axes. In Sec. III, two-photon polarization matrices are derived for each Landau state. The correlations between the polarizations of the photons are considered in Sec. IV. Finally, the conclusions are given in Sec. V.

\section{Angular distributions}

In the center of mass reference frame, the wave function in momentum space of TP depends only on the direction of the relative momentum $\mathbf{n}\left(\theta ,\varphi \right)$. The Landau TPSS states \cite{Jacob,LLIV} in Table II are given in terms  of the two-photon spherical helicity states  $\left\vert JM;\lambda \lambda ^{\prime }\right\rangle $ defined by
\begin{equation}
\left\vert JM;\lambda \lambda ^{\prime }\right\rangle =\sqrt{\frac{2J+1}{%
4\pi }}D_{\lambda -\lambda ^{\prime },M}^{\left( J\right) }\left( \mathbf{n}%
\right) \left\vert \lambda ,\lambda ^{\prime }\right\rangle,  \tag{1}
\end{equation}%
which are eigenfunctions of the angular momentum, the projection of the angular momentum, and the total helicity of the photon pair. In (1), $D_{\lambda M}^{( J)}$ is the Wigner function and $|\lambda,\lambda^{\prime}\rangle \equiv | \textbf{n};\lambda^{\prime} \otimes |-\textbf{n} ;\lambda^{\prime}\rangle$; $(\lambda,\lambda ^{\prime}=\pm 1)$ are eigenfunctions of the total helicity with eigenvalue $\lambda -\lambda ^{\prime }$. In succeeding calculations, we use the expression
\begin{equation}
D_{\lambda M}^{\left( J\right) }\left( \mathbf{n}\right) =e^{iM\varphi
}d_{\lambda M}^{\left( J\right) }\left( \theta \right)  \tag{2}
\end{equation}%
for the Wigner function \cite{LLIV}. In (1), $J=0, 2, 3\cdots $ and the absence of $J=1$ is a direct consequence of the transversality of the two-photon system in the center of mass reference frame \cite{LLIV}. Each TPSS in Table II is either the sum or the difference of two-photon spherical helicity states given in (1) with opposite total helicity.

\begin{table*}[ht]
\caption{All possible two-photon spherical states classified by angular momentum $J$ and parity $P$} 
\centering 
\begin{tabular}{ccc} 
\hline\hline 
$J$  &\ $P = +1$ &\ $P=-1$  \\ [0.5ex] 
\hline 
0 & $ \frac{1}{\sqrt{2}}(|00;11\rangle + |00;-1-1\rangle)$ & $\frac{1}{\sqrt{2}}(|00;11\rangle - |00;-1-1\rangle)$  \\ 
1 & $-$ & $-$ \\
 & a. \hspace{.01in}$\frac{1}{\sqrt{2}}(|JM;11\rangle + |JM;-1-1\rangle)$ &   \\
 2$k$& &$\frac{1}{\sqrt{2}}(|JM;11\rangle - |JM;-1-1\rangle)$  \\
 & b. \hspace{.01in} $\frac{1}{\sqrt{2}}(|JM;1-1\rangle + |JM;-11\rangle)$ &  \\
2$k$+1 &  $\frac{1}{\sqrt{2}}(|JM;1-1\rangle - |JM;-11\rangle)$ & $-$  \\ [1ex] 
\hline 
\end{tabular}
\label{table:Table I} 
\end{table*}

There are two axes in these problems, viz., the quantization axis $Z$ that is defined by the source of radiation and $\mathbf{n}$. Yang \cite{Yang} uses plane waves propagating along and opposite the direction of $\mathbf{n}$. If the quantization axis is coincident with $\mathbf{n}$, then%
\begin{equation}
d_{\lambda M}^{\left( J\right) }\left( 0\right) =\delta _{\lambda M}  \tag{3}
\end{equation}
and our Table II corresponds to Yang's table.

The angular distribution of the TP radiation is actually the density of registration of photon pairs at given direction $\mathbf{n}$, which is defined by the square of the modulus of the wave functions in Table II. Using (1) for the two-photon spherical helicity states and (2) for the Wigner function, we obtain the angular distribution for Landau's TPSS
\begin{equation}
w_{\left\vert \Lambda \right\vert }^{JM}=\frac{2J+1}{8\pi }\left[ d_{\Lambda
M}^{\left( J\right) 2}+d_{-\Lambda M}^{\left( J\right) 2}\right],   \tag{4}
\end{equation}%
which depends on the modulus of the total helicity $\Lambda =\lambda -\lambda^{\prime }$, viz., $\left\vert \Lambda \right\vert =0, 2$. The absence of interference terms in (4) is a direct consequence that the states listed in Table II are linear combinations of orthogonal, helicity states. Using the Clebsch-Gordan series representation of two Wigner functions product \cite{Varsh}, we obtain the final expression for the angular distribution (4)
\begin{equation}
w_{m}^{JM}\left( \theta \right) =A_{JM}\sum\limits_{n=0}^{J}\left(4n+1\right) \Big{(}
\begin{array}{ccc}
J & J & 2n \\
m & -m & 0%
\end{array}%
\Big{)}
\Big{(}
\begin{array}{ccc}
J & J & 2n \\
M & -M & 0%
\end{array}%
\Big{)} P_{2n}\left( \cos \theta \right), \hspace{0.2in} (m=0, 2)  \tag{5}
\end{equation}%
where $A_{JM}=\left( -1\right) ^{M}(2J+1)/(4\pi)$.

Table III indicates the angular distributions of the TPSS in accordance with Landau's Table I.
\begin{table}[th]
\caption{Angular distribution of the two-photon spherical states}
\label{table:two}
\centering 
\begin{tabular}{ccc}
\hline\hline
$J$ & \ $P = +1$ & \ $P=-1$ \\[0.5ex] \hline
0 & $\frac{1}{4\pi}$ & $\frac{1}{4\pi}$ \\
1 & $-$ & $-$ \\
 & a.\hspace{.04in}$w_{0}^{JM}$ &  \\
 2$k$ &  &$w_{0}^{JM}$\\
  & b.\hspace{.04in}$w_{2}^{JM}$ &  \\
 2$k$+1  & $w_{2}^{JM}$ & $-$ \\[1ex] \hline
\end{tabular}
\end{table}

\section{Polarization matrices}

We introduce normalized two-photon polarization matrix $\hat{\rho}$ \cite{F1,vem} for each TPSS of Table II via the pure spherical state density operator
\begin{equation}
\left\vert PJM\right\rangle \left\langle PJM\right\vert =w_{\left\vert \Lambda \right\vert }^{JM}\hat{\rho},  \tag{6}
\end{equation}
where $w_{\left\vert \Lambda \right\vert }^{JM}$ is the angular distribution of the radiation that is coincident with the trace of the density operator. Substituting the expression of every state in Table II into (6) and using the factorization via a Kronecker product of the Pauli matrices $\hat{\sigma}_{i}\otimes \hat{\sigma}_{j}^{\prime }$ \cite{Blum},  we obtain the entries in Table IV for the normalized two-photon polarization matrices
\begin{table}[th]
\caption{Two-photon polarization matrices of Landau's states}
\label{table:two}
\centering 
\begin{tabular}{ccc}
\hline\hline
$J$ & \ $P = +1$ & \ $P=-1$ \\[0.5ex] \hline
0 & $\hat{\rho}^{+}$ & $\hat{\rho}^{-}$ \\
1 & $-$ & $-$ \\
 & a. $\hspace{.04in}$$\hat{\rho}^{+}$ & {--} \\
 2$k$ &  &$\hat{\rho}^{-}$\\
  & b. \hspace{.04in}$\hat{\rho}^{JM}_{e}$ &  \\
 2$k$+1  & $\hat{\rho}_{o}^{JM}$ & $-$ \\[1ex] \hline
\end{tabular}
\end{table}

In Table IV, the two-photon polarization density matrices $\hat{\rho}^{\pm }$
\begin{equation}
\hat{\rho}^{\pm }=\frac{1}{4}\left\{ \hat{I}\otimes I^{\prime }+\hat{\sigma}%
_{3}\otimes \hat{\sigma}_{3}^{\prime }\pm \left( \hat{\sigma}_{1}\otimes
\hat{\sigma}_{1}^{\prime }-\hat{\sigma}_{2}\otimes \hat{\sigma}_{2}^{\prime
}\right) \right\}   \tag{7}
\end{equation}%
do not depend on $J$ and $M$ and are associated with even and odd parities of TPSS. They describe fully entangled states of polarization of the photon pair, i.e., the photons are not polarized individually but the two-photon system is polarized completely \cite{vem,F2}). Now $\hat{\rho}_{e,o}^{JM}$

\begin{equation}
\hat{\rho}_{e,o}^{JM}  = \frac{1}{4} \Big{\{} \hat{I}\otimes I^{\prime }+\xi^{JM}\left( \hat{\sigma}_{3}\otimes \hat{I}^{\prime }-\hat{I}\otimes \hat{\sigma}_{3}^{\prime }\right)\\
 -\hat{\sigma}_{3}\otimes \hat{\sigma}_{3}^{\prime }
 \pm \zeta ^{JM}(\hat{\sigma}_{1}\otimes \hat{\sigma}_{1}^{\prime}+\hat{\sigma}_{2}\otimes \hat{\sigma}_{2}^{\prime} \Big{\}} \tag{8}
\end{equation}
depends on $J$ and $M$ and are associated with TPSS with even ($e$) and odd ($o$) angular momentum $J$ $\left( J\geq 2\right)$. In (8), the polarization parameters $\xi ^{JM}$ and $\zeta ^{JM}$ are defined by \begin{equation}
\xi ^{JM}=\frac{2J+1}{8\pi w_{2}}\left[ d_{2M}^{\left( J\right)
2}-d_{-2M}^{\left( J\right) 2}\right] \text{ }  \tag{9.a}
\end{equation}%
\begin{equation}
\zeta ^{JM}=\frac{2J+1}{4\pi w_{2}}d_{2M}^{\left( J\right) }d_{-2M}^{\left(
J\right) }  \tag{9.b}
\end{equation}%
that depend on the polar angle $\theta $ formed by the quantization axis $Z$ and $\mathbf{n}$. The matrix (8) describes correlated states of the individually, partially circularly polarized photons wherein
the system of two photons is polarized completely \cite{F1, vem}.

In the case of photons flying parallel to the quantization axes ($\theta =0$), the matrices $\hat{\rho}_{e,o}^{JM}$ describe uncorrelated states of the two fully, circularly polarized photons. Actually, using identity (3), we find $\xi ^{JM}=\delta _{M,2}-\delta _{M,-2}$ and $\zeta ^{JM}=0$ for the polarization parameters (9). Accordingly, the matrices (8) become
\begin{equation*}
\hat{\rho}_{e,o}^{JM}=\left\{
\begin{array}{c}
\left(\hat{I}+\hat{\sigma}_{3}\right) \otimes \left( \hat{I}^{\prime }-\hat{\sigma}_{3}^{\prime }\right) \text{ } \hspace{.15in} (M=2) \\
\left(\hat{I}-\hat{\sigma}_{3}\right) \otimes \left( \hat{I}^{\prime }+\hat{\sigma}_{3}^{\prime }\right). \text{ } \hspace{.15in}(M=-2)%
\end{array}%
\right.
\end{equation*}

The next simple case for expression (8) is the case for photons flying perpendicular to the quantization axes ($\theta =\pi/2$). Using the symmetry property of the $d_{M^{\prime }M}^{\left( J\right) }$ functions%
\begin{equation*}
d_{M^{\prime }M}^{\left( J\right) }\left( \pi -\theta \right) =\left(
-1\right) ^{J-M}d_{-M^{\prime }M}^{\left( J\right) }\left( \theta \right),
\end{equation*}%
 we find $d_{2M}^{\left( J\right) }\left( \pi/2\right) =\left( -1\right) ^{J-M}d_{-2M}^{\left( J\right) }\left( \pi/2\right) $ and, with the aid of (4) and (9), we obtian
\begin{equation*}
\xi ^{JM}=0,\text{ }\zeta ^{JM}=\left( -1\right) ^{J-M}
\end{equation*}%
then the polarization matrices (8) reduces to

\begin{equation}
\hat{\rho}_{e}^{JM} = \hat{\rho}_{o}^{JM} = \frac{1}{4}\big{\{}\hat{I}\otimes I^{\prime}-\hat{\sigma}_{3}\otimes \hat{\sigma}_{3}^{\prime}
 +(-1)^{M}(\hat{\sigma}_{1}\otimes \hat{\sigma}_{1}^{\prime}+\hat{\sigma}_{2}\otimes \hat{\sigma}_{2}^{\prime})\big{\}},\tag{10}
\end{equation}
i.e., at the angle $\theta =\pi /2$ the polarization matrix of the two-photon system depends only on the projection $M$ of the angular momentum  and again we come to a fully entangled polarization state of the individual photons when the photon pair is completely polarized.

Finally, using the Clebsch-Gordan series representation of two Wigner functions product \cite{Varsh}, we obtain the expressions for the polarization parameters $\xi^{JM}$ and $\zeta^{JM}$ of (9) in the form
\begin{equation}
\xi ^{JM}\left( \theta \right) =\frac{A_{JM}}{w_{2}}\sum\limits_{n=1}^{J}\left( 4n-1\right)
\Big{(}\begin{array}{ccc}
J & J & 2n-1 \\
2 & -2 & 0%
\end{array}\Big{)}%
\Big{(}\begin{array}{ccc}
J & J & 2n-1 \\
M & -M & 0%
\end{array}\Big{)}%
 P_{2n-1}\left( \cos \theta \right).   \tag{11.a}
\end{equation}%

\begin{equation}
\zeta ^{JM}\left( \theta \right) =\frac{A_{JM}}{w_{2}}\sum\limits_{n=2}^{J}%
\left( 4n+1\right) \left[ \frac{\left( 2n-4\right) !}{\left( 2n+4\right) !}%
\right] ^{1/2}\Big{(}
\begin{array}{ccc}
J & J & 2n \\
2 & 2 & -4%
\end{array}%
\Big{)}
\Big{(}
\begin{array}{ccc}
J & J & 2n \\
M & -M & 0%
\end{array}%
\Big{)} P_{2n}\left( \cos \theta \right).  \tag{11.b}
\end{equation}

\section{Polarization correlations}

The angular distributions (5) of the two-photon radiation contains information on the quantum numbers $J$ and $M$. Information related to the parity $P$ of TPSS is determined by linear polarization correlation measurements \cite{Yang}.

Photon polarization are measured by means of polarization analyzers whose efficiency depends on the value of the helicity. The efficiency of the polarization analyzer is determined by the efficiency matrix with
orientation vector $\boldsymbol{\eta }$ ($\eta =1$) having the form of a Stokes matrix, viz.,
\begin{equation}
\hat{\epsilon}=\frac{1}{2}(\hat{I}+\text{\textbf{$\eta $}}\cdot \mathbf{\hat{%
\sigma}}),  \tag{12}
\end{equation}%
which acts in the helicity Hilbert subspace reducing the mixed quantum state of polarization into a pure quantum state \cite{RB}.

The two-photon polarization is measured by two polarization analyzers $\hat{\epsilon}$ and $\hat{\epsilon}^{\prime}$ measuring the helicity states of the individual photons propagating along $\mathbf{n}$ and $\mathbf{n}^{\prime}$, respectively. The result of the measurement is defined as the probability of the reduction of the two-photon polarization matrix $\hat{\rho}$ into the pure quantum state of two uncorrelated photons described by $\hat{\epsilon}\otimes \hat{\epsilon}^{\prime }$, i.e.,

\begin{equation}
W(\mathbf{\eta ,\eta }^{\prime })=\textup{Tr}(\hat{\rho}\hat{\epsilon}\otimes \hat{%
\epsilon}^{\prime }).  \tag{13}
\end{equation}

In other words, $W(\mathbf{\eta ,\eta }^{\prime })$ is the probability of registering a photon in the definite helicity state $\left\vert \mathbf{n;}\lambda \right\rangle $ in coincidence with the other photon in the helicity state $\left\vert \mathbf{n}^{\prime }\mathbf{;}\lambda ^{\prime}\right\rangle $. Measurement of the probability $W(\mathbf{\eta,\eta }^{\prime })$ involves definite orientations and the kind of analyzers used. These measurements characterize the quantum state of the photon pair in an unambiguous way \cite{vem, ma}.

The efficiency matrix $\hat{\epsilon}$ of an analyzer measuring the linear polarization $\mathbf{\eta }$ having oriented angle $\psi $ relative to axes $x$ for the photon propagation direction $\mathbf{n}$ is given by \cite{Blum}
\begin{equation}
\hat{\epsilon}=\frac{1}{2}\left( \hat{I}-\cos(2\psi) \hat{\sigma}_{1}-\sin(2\psi) \hat{\sigma}_{2}\right).  \tag{14.a}
\end{equation}%
The efficiency matrix $\hat{\epsilon}^{\prime }$ $\left( \mathbf{\eta }^{\prime }\right)$ for the photon propagating along $\mathbf{-n}$ is similarly
\begin{equation}
\hat{\epsilon}^{\prime }=\frac{1}{2}\left( \hat{I}^{\prime }-\cos(2\psi^{\prime}) \hat{\sigma}_{1}^{\prime }+\sin(2\psi^{\prime }) \hat{\sigma}_{2}^{\prime }\right).  \tag{14.b}
\end{equation}%
On substituting (14) in (13), we find for the two-photon polarization density matrices $\hat{\rho}^{\pm }$ (7)
\begin{equation}
W^{\pm }(\psi ,\psi ^{\prime })=\textup{Tr}(\hat{\rho}^{\pm }\hat{\epsilon}\otimes
\hat{\epsilon}^{\prime })=\frac{1}{2}\left\{
\begin{array}{c}
\cos ^{2}\left( \psi -\psi ^{\prime }\right)  \\
\sin ^{2}\left( \psi -\psi ^{\prime }\right).
\end{array}%
\right . \tag{15}
\end{equation}%

Probability (15) indicates that in the state with positive parity, the photons are polarized predominately with parallel linear polarizations. On the other hand, with negative parity, the photons have linear polarization that are predominately perpendicular to each other. In addition, the photon polarization correlations do not have angular dependence, viz., photons are invariant relative to $Z$ to spatial rotations.

These type of correlations were mentioned by Yang \cite{Yang}. In addition, $W^{+}(\psi ,\psi ^{\prime })$ is the well-known $\left( \mathbf{ee}^{\prime }\right)^{2}$ polarization correlation for the Goeppert-Mayer's two-photon emission \cite{GM, Br-Tel}. Owing to momentum and parity conservation for the 2S-1S atomic transition, we have the emission of the two photons in the $\left\vert 00+1\right\rangle $\ Landau state.

For the polarization density matrices $\hat{\rho}_{e,o}^{JM}$ (8) we find, in an analogous way,
\begin{equation}
W_{e,o}(\psi ,\psi ^{\prime })=\textup{Tr}(\hat{\rho}_{e,o}\hat{\epsilon}\otimes \hat{%
\epsilon}^{\prime })=\frac{1}{4}\left\{
\begin{array}{c}
1+\zeta ^{JM}\cos[2\left(\psi +\psi ^{\prime}\right)]  \\
1-\zeta ^{JM}\cos[ 2\left( \psi +\psi ^{\prime }\right)],
\end{array}%
\right.   \tag{16}
\end{equation}
which clearly depends on $\theta$.

It should be remarked that in (16), the angles $\psi$ and $\psi^{\prime}$ in (14) are defined in terms of the $x$ axis. In addition, the direction of \textbf{n} is arbitrary relative to the quantization axis $Z$ and so one can take the direction of $x$ as being normal to the plane (\textbf{n},\textbf{Z)}. Consider the choice $\psi^{\prime} = 0$, that is, the analyzer $\hat{\epsilon}^{\prime}$ is orientated along the $x$-axis, then (16) becomes
\begin{equation}
W_{e,o}(\psi ,\psi ^{\prime })=\textup{Tr}(\hat{\rho}_{e,o}\hat{\epsilon}\otimes \hat{%
\epsilon}^{\prime })=\frac{1}{4}\left\{
\begin{array}{c}
1+\zeta ^{JM}\cos(2\psi)  \\
1-\zeta ^{JM}\cos(2\psi).
\end{array}%
\right.   \tag{17}
\end{equation}
If \textbf{n} and $Z$ are parallel, that is, $\theta=0$, then the photon correlation disappears since $\zeta^{JM}=0$.

Expression (17) for the correlation is further simplified for the case $\theta= \pi/2$ (see (10)), we then have
\begin{equation}
W_{e}(\psi ,\psi ^{\prime })=W_{o}(\psi ,\psi ^{\prime }) =\frac{1}{2}\left\{
\begin{array}{c}
\cos^2\psi  \hspace{.1in} (M=even) \\
\sin^2\psi  \hspace{.1in} (M=odd).
\end{array}%
\right.   \tag{18}
\end{equation}
Accordingly, if one of the linear analyzers is oriented normal to the (\textbf{n},$Z$) plane, then for M even the photons are polarized predominantly parallel to the plane while for M odd, the polarization is predominantly perpendicular to the plane.

In closing, we note that consideration of the correlation between circularly polarized photons does not provide any information about the parity of the two-photon state.

\section{Conclusion}

We have analyzed in detail the angular-polarization properties of the Landau two-photon spherical states in momentum space in the center of mass reference frame of the two-photon system. The angular distributions do not depend on the parity and, for fixed values of $J$ and $M$, are defined by two different functions of the polar angle between the relative momentum and the quantization axes. The two-photon polarization density matrices are derived for each values of $J$, $M$, and $P$. The linear polarization correlations of individual photons are analyzed in details. We find, besides the ordinary correlation laws given in terms of $sin$ and $cos$ of the angle between the orientation of the analyzers for $J\geq 2$, also the correlations in terms of $sin$ and $cos$ of the sum of the orientation angles of analyzers.\\


\begin{thebibliography}{99}
\bibitem{Br-Tel} G.Breit and E.Teller, Metastability of hydrogen and helium levels, Astrophys. J. \textbf{91} (1940) 215--239.
\bibitem{GM} M. Goeppert-Mayer, \"{U}ber Elementarakte mit zwei Quantenspr \"{u}ngen, Ann. Phys. (Leipzig) \textbf{9} (1931) 273--295 .
\bibitem{Sun} J. Chluba and R.A. Sunyaev, Two-photon transitions in hydrogen and cosmological recombination A \& A \textbf{480,} (2008) 629--645.
\bibitem{Peeb} P.J.E. Peebles, \textit{Principles of physical cosmology} (Princeton University Press, Princeton, NJ, 2020).
\bibitem{Asp} A. Aspect, J. Dalibard, and G. Roger, Experimental test of Bell's inequalities using time-varying analyzers, Phys. Rev. Lett. \textbf{49} (1982) 1804--1807.
\bibitem{Klei} H. Kleinpoppen, A.J. Duncan  \emph{et al.}, Coherence and polarization analtsis of the two-photon radiation of metastable atomic hydrogen, Phys. Script. T \textbf{72} (1997) 7--17.
\bibitem{Gis} W. Tittel, J. Brendel, \emph{et al.}, Violation of Bell inequalities by photons more than 10 km apart, Phys. Rev. Lett. \textbf{81} (1998) 3563--3567.
\bibitem{Sch} A. Sch\"{a}fer, G. Soff, \emph{et al.}, Prospects for an atomic parity-violation experiment in $U^{90+}$, Phys. Rev. A \textbf{40} (1989) 7362--7365.
\bibitem{Frat} F. Fratini, S. Trotsenko, \emph{ et al.}, Photon-photon polarization correlations as a tool for studying parity nonconservation in heliumlike uranium, Phys. Rev. A \textbf{83} (2011) 052505.
\bibitem{Land} L.D. Landau, On the momentum of a system of two photons (in Russian), Dokl. Akad. Nauk USSR, \textbf{60} (1948) 207--209.
\bibitem{Yang} C. N. Yang, Selection rules for the dematerialization of a particle into two photons, Phys. Rev. \textbf{77} (1950)  242--245.
\bibitem{DBD83} D. DeMille, D. Budker, N. Derr, and E. Deveney, Search for Exchange-Antisymmetric Two-photon States, Phys. Rev. Lett. \textbf{83} (1999) 3978.
\bibitem{KEB09} M.G. Kozlov, D. English, and D. Budker, Symmetry-suppressed two-photon transitions induced by hyperfine interactions and magnetic fields, Phys. Rev. A \textbf{80} (2009) 042504.
\bibitem{ZSL15} T. Zalialiutdinov, D. Solovyev, L. Labzowsky, and G. Plunien, Exclusion principle for photons: Spin-statistic selection rules for multiphoton transitions in atomic systems, Phys. Rev. A \textbf{91} (2015) 033417.
\bibitem{ZSL17} T. Zalialiutdinov, D. Solovyev, L. Labzowsky, Generalized spin-statistic selection rules for atomic transitions with arbitrary number of equivalent photons, The European Physical Journal Special Topics  \textbf{226} (2017) 2837-2842.
\bibitem{Jacob} M. Jacob and G.C. Wick, On the General Theory of Collisions for Particles with Spin, Ann. Phys. (N.Y.) \textbf{7} (1959) 404--428.
\bibitem{LLIV} V.B. Berestetskii, E.M. Lifshitz, and L.P. Pitaevskii, \textit{Quantum Electrodynamics} (Pergamon Press, Oxford, 2010).
\bibitem{Jack} J.D. Jackson, \textit{Classical electrodynamics} (John Wiley \& Sons, Inc., 1998).
\bibitem{Akh} A.I. Akhiezer and V.B. Berestetskii, \textit{Quantum electrodynamics} (John Wiley \& Sons, Inc., 1965).
\bibitem{Varsh} D. A. Varshalovich, A. N. Moskalev, and V. K. Khersonskii, \textit{Quantum Theory of Angular Momentum} (World Scientific, Singapore, 1988).
\bibitem{F1} U. Fano, Description of states in quantum mechanics by density matrix and operator techniques, Rev. Mod. Phys. \textbf{29} (1957) 74--94.
\bibitem{vem} V.E. Mkrtchian and V.O. Chaltykian, Polarization states of the two-photon system, Opt. Comm. \textbf{63} (1987) 239--242.
\bibitem{Blum} K. Blum, \textit{Density matrix theory and applications} (Springer-Verlag, Berlin, Heidelberg, 2012).
\bibitem{F2} U. Fano, Pairs of two-level systems, Rev. Mod. Phys. \textbf{55} (1983) 855--874 .
\bibitem{RB} R. Balian, On the principles of quantum mechanics and the reduction of the wave packet, Am. J. Phys. \textbf{57} (1989) 1019--1027.
\bibitem{ma} M. Alexanian and V.E. Mkrtchian, Quantum entropy and polarization measurement of the two-photon system, Phys. Rev. A \textbf{97} (2018) 022326(8).


\end{thebibliography}
\end{document}